\documentstyle[12pt,worldsci]{article}
\def \lsim {~\mbox{${}^< \hspace*{-7pt} _\sim$}~}
\def \gsim {~\mbox{${}^> \hspace*{-7pt} _\sim$}~}

\def\preRate{\Bigl[ N \frac{\rho_\chi}{m_\chi} \Bigr]}
\def\Rate0{\Bigl[ N \frac{\rho_\chi}{m_\chi} \sigma(0)
                  \langle v \rangle \Bigr]}
\def\dsdq2{\frac{d\sigma}{dq^2}}

\def\and{\mbox{~~~and~~~}}
\newcommand{\be}[1]{\begin{equation} \label{(#1)}}
\newcommand{\ee}{\end{equation}}
\newcommand{\ba}[1]{\begin{eqnarray} \label{(#1)}}
\newcommand{\ea}{\end{eqnarray}}
\newcommand{\nn}{\nonumber}
\newcommand{\rf}[1]{(\ref{(#1)})}

\begin{document}
\setlength{\unitlength}{1in}

\begin{center}
{\bf Expectations for Supersymmetric Dark Matter\\
     Searches Underground.}

\bigskip
{V.A. Bednyakov$^1$, H.V. Klapdor-Kleingrothaus$^2$, S.G. Kovalenko$^1$
\bigskip

$^1${\it Joint Institute for Nuclear Research, Dubna, Russia}
\bigskip

$^2${\it
Max-Plank-Institut f\"{u}r Kernphysik, D-6900, Heidelberg,
Germany}
}\\[0.5cm]

{\sf presented by S.G. Kovalenko}
\end{center}
\bigskip

\begin{center}
{\bf Abstract}
\bigskip

\begin{minipage}[c]{120mm}
        {}\hspace*{0.5cm}
        We consider the neutralino as a dominant Dark Matter
        particle in the galactic halo and investigate some general issues
        of direct DM searches via elastic neutralino-nucleus scattering.
        On the basis of conventional assumptions about the nuclear and
        nucleon structure we analyse constraints on SUSY model
        parameter space accessible by the direct DM searches.
        This analysis shows that DM detectors fall into
        the three different
        categories with respect to their sensitivity to different
        groups of the SUSY model parameters.

	{}\hspace*{0.5cm}
        We calculate the event rate for various experimentally interesting
        isotopes within the Minimal Supersymmetric Standard Model (MSSM)
        taking into account the known accelerator and cosmological constraints
        on the MSSM parameter space.

	{}\hspace*{0.5cm}
        We investigate the role of nuclear spin in elastic neutralino-nucleus
        scattering. It is found that the contribution of the spin-dependent
        interaction to this process is subdominant for
        nuclei with atomic weights $A\geq 50$.
\end{minipage}
\end{center}

\bigskip
\section{Introduction}
        Analysis of the data on distribution and motion of astronomical
        objects within our galaxy and far beyond indicates presence
        of a large amount of non-luminous dark matter
        (for review see\cite{Relic}).
        According to estimations, dark matter (DM) may constitute more
        than $90\%$ of the total mass of the universe if a mass density
        $\rho$ of the universe is  assumed to be close to the critical value
        $ \rho_{c} = 3 H^2/8\pi G_N$ (H is the Hubble constant and
        $G_N$ is the Newtonian gravitational constant).
        The exact equality $\Omega=\rho/\rho_{c}=1$, corresponding to
        a flat universe,
        is supported by naturalness arguments and by inflation scenarios.
        The theory of primordial nucleosynthesis restricts the amount
        of baryonic matter in the universe to $\sim$10\%.
        Thus a dominant component of DM is non-baryonic.
        The recent data by the COBE satellite\cite{COBE} on  anisotropy
        in the cosmic background radiation and the theory of the
        formation of large scale structures
        of the universe lead to the conclusion that non-baryonic DM itself
        consists of a dominant ($70\%$) "cold" DM (CDM) and a smaller
        ($30\%$) "hot" DM (HDM) component\cite{Taylor,Davis}{}.

        A possible HDM candidate is the massive neutrino.
        The SUSY model neutralino ($\chi$) is currently a favorable
        candidate for CDM.
        This is a Majorana ($\chi^{c}=\chi$) particle with spin $1/2$
	predicted by supersymmetric (SUSY) models.

        There are four neutralinos in the minimal supersymmetric extension
	of the standard model (MSSM) (see\cite{Haber}).
        They are a mixture of gauginos ($\tilde{W}_{3}, \tilde{B})$
        and Higgsinos ($\tilde{H}_{1,2}$) being SUSY partners of gauge
        ($W_{3}, B$) and Higgs ($H_{1,2}$) bosons.
        The DM neutralino $\chi$ is assumed to be the lightest
	supersymmetric particle (LSP)
        and therefore is stable in SUSY models with $R$-parity conservation.

        In our galaxy most of the mass is expected to be in form of
	a spherical dark halo.
        Microlensing searches have discovered far more MACHOs
        in the disk than in the halo of the Galaxy. The data obtained
        from these searches is consistent with the fraction of MACHOs
        in the halo dark matter\cite{GGT} less than 30\%.
        Thus, most of the halo of the galaxy must be non baryonic
        cold dark matter.

	In the galactic halo neutralinos are assumed
	to be  Maxwellian distributed in
        velocities with a mean velocity in the earth frame
        $v\approx$ 320 km/sec\cite{Gould}{}.
	Their mass density in the Solar system
	is expected to be about
	\mbox{$\rho \approx 0.3 $ GeV $\cdot$ cm$^{-3}$.}
        Therefore, at the earth surface neutralinos might produce a
        substantial flux ($\Phi = \rho\cdot v/M$) of
        $\Phi>10^{7}$cm$^{-2}$ sec$^{-1}$
        for a particle mass of M$\sim$ 1 GeV.
        In view of this one may hope to detect DM particles directly,
	for instance
        through elastic scattering from nuclei inside terrestrial detector.

        The problem of direct detection of the DM neutralino $\chi$  via
        elastic scattering off nuclei has attracted considerable efforts
        during the last decade and remains a field of great experimental
        and theoretical
        activity\cite{Witt,Griest,Gelm,Bot,EF,Drees,BKK,BKK1,TAUP}{}.

        In this report we address the questions concerning
        prospects for the direct detection of the supersymmetric
        Dark Matter with the current and the near future detectors.

        In Sec. 2 we consider general issues of such experiments for
        exploration of the SUSY model parameter space and classify
        possible DM detectors with respect to their sensitivity
        to different domains of this parameter space.
        We propose and discuss special criterion for assessing
        an isotope as a target material for a DM detector.
        In discussing general expectations for DM detection experiments,
        we avoid the use of specific nuclear and nucleon structure models,
        but rather base our consideration on the known experimental data about
        nuclei and nucleon properties.

        In Sec. 3 we discuss predictions for the DM detection
        event rate obtained
        in the framework of the MSSM. We undertake a systematic exploration
        of a broad domain of the MSSM parameter space
        restricted by the well known accelerator constraints
        and by the cosmological bounds on neutralino relic abundance in
        the universe. The effect of a non-zero threshold energy
        of a realistic DM detector is analyzed.

        Sec. 4 is devoted to the role of nuclear spin in
        the DM neutralino detection.
        In general, the event rate $R$ for elastic $\chi$-nucleus
        scattering contains contributions from the spin-dependent ($R_{sd}$)
        and spin-independent ($R_{si}$) neutralino-nucleus
	interactions: $R=R_{sd}+R_{si}$.
        We have found that the $R_{si}$ contribution dominates in the total
        event rate $R$ for nuclei with atomic weight $A>50$ in the
        region of the MSSM parameter space where
        $R =R_{sd}+R_{si} > 0.01\frac{ events }{  kg \cdot day }$.
        The lower bound  $0.01 \frac{ events }{  kg \cdot day }$
	seems to be far below the sensitivity of realistic present
	and near future DM detectors. Therefore one can ignore
        the region where $R < 0.01 \frac{ events }{  kg \cdot day }$ as
        invisible for these detectors.

        In view of this result we do {\it not\ } expect crucial
	dependence of the DM event rate on the nuclear spin for detectors
	with target nuclei having an atomic weight larger than $50$.
        In other words, we expect essentially {\em equal chances}
	for J~=~0 and J~$\neq$~0 detectors to discover DM events.

        In particular, this conclusion supports the idea that presently
        operating $\beta\beta$-detectors with spinless nuclear target
        material have good prospects for DM neutralino search.

        Sec. 5 gives a conclusion.

\section{General Properties of the Neutralino-Nucleus Interactions}

        A DM event is elastic scattering of a DM neutralino from
	a target nucleus
	producing a nuclear recoil which can be detected by a detector.
        The corresponding event rate depends on the distribution of
        the DM neutralinos in the solar vicinity and
        the cross section $\sigma_{el}(\chi A)$ of  neutralino-nucleus
        elastic scattering.
        In order to calculate $\sigma_{el}(\chi A)$  one should specify
        neutralino-quark interactions.
        The relevant low-energy effective Lagrangian can be written
        in a general form as
\be{Lagr} 
  L_{eff} = \sum_{q}^{}\left( {\cal A}_{q}\cdot
		\bar\chi\gamma_\mu\gamma_5\chi\cdot
                \bar q\gamma^\mu\gamma_5 q +
    \frac{m_q}{M_{W}} \cdot{\cal C}_{q}\cdot\bar\chi\chi\cdot\bar q q\right)
      \ +\ O\left(\frac{1}{m_{\tilde q}^4}\right),
\ee
        where terms with the vector and pseudoscalar quark currents are
        omitted being negligible in the case of the non-relativistic
        DM neutralino with typical velocities $v_\chi\approx 10^{-3} c$.

        In the Lagrangian \rf{Lagr} we also neglect terms which appear in
	supersymmetric models at the order of $1/m_{\tilde q}^4$
	and higher,  where $m_{\tilde q}$ is
	the mass of the scalar superpartner $\tilde q$
	of the quark $q$. These terms, as recently pointed out
	by Drees and Nojiri\cite{Drees}{}, are potentially
	important in the spin-independent neutralino-nucleon scattering,
	especially in domains of the MSSM  parameter space where
	$m_{\tilde q}$ is close to the neutralino mass $M_{\chi}$.
	Below we adopt the approximate treatment of these terms
        proposed in Ref.\cite{Drees} which allows "effectively"
        absorbing them into the coefficients ${\cal C}_q$ in a wide region
	of the SUSY model parameter space.

	The coefficients ${\cal A}_{q}, \ {\cal C}_{q}$ depend on
        the specific SUSY model and will be considered
	in the next section.

	Here we survey general properties of neutralino-nucleus
        ($\chi$-$A$) scattering following from the Lagrangian \rf{Lagr}.

        A general representation of the differential cross section
        of neutralino-nucleus scattering can be given in terms of
        three spin-dependent ${\cal  F}_{i}(q^2)$ and
        one spin-independent ${\cal F}_{S}(q^2)$ form factors
        as follows\cite{EV}
\be{cs}
\dsdq2(v,q^2)=\frac{8 G_F}{v^2} \left(
a_0^2\cdot {\cal F}_{00}^2(q^2) +
a_0 a_1 \cdot {\cal F}_{10}^2(q^2) +
a_1^2\cdot {\cal F}_{11}^2(q^2)
+ c_0^2\cdot A^2\ {\cal F}_{S}^2(q^2)
\right).
\ee
	The last term corresponding to the spin-independent scalar interaction
	gains coherent enhancement $A^2$  ($A$ is the atomic weight of
	the nucleus in the reaction).
	Coefficients $a_i, c_0$ do not depend on nuclear structure
	and relate to the parameters ${\cal A}_q, {\cal C}_q$ of the effective
	Lagrangian \rf{Lagr} and to the parameters $\Delta q, f_s, \hat{f}$
	characterizing nucleon structure. One has the relationships
\ba{rel1}
a_0 &=& \left({\cal A}_u + {\cal A}_d\right)
      \left(\Delta u + \Delta d \right) + 2 \Delta s {\cal A}_s,\ \ \ \
a_1 = \left({\cal A}_u - {\cal A}_d\right)
      \left(\Delta u - \Delta d \right),\\ \nn
c_0  &=& \hat f \frac{m_u {\cal C}_{u}
        + m_d {\cal C}_{d}}{m_u + m_d}
+ f_{s} {\cal C}_{s}  + \frac{2}{27}(1- f_{s} - \hat f)({\cal C}_{c}
        + {\cal C}_{b} + {\cal C}_{t})
\ea
	Here  $\Delta q^{p(n)}$ are
        the fractions of the proton(neutron) spin carried by the quark $q$.
        The standard definition is
\be{Spin} 
   <p(n)|\bar q\gamma^\mu\gamma_5 q|p(n)> = 2 S_{p(n)}^{\mu} \Delta q^{p(n)},
\ee
      	where $S_{p(n)}^{\mu}=(0,\vec{S}_{p(n)})$ is the 4-spin of the nucleon.
        The parameters $\Delta q^{p(n)}$ can be extracted
	from data on polarized nucleon
        structure functions\cite{EMC,SMC2} and hyperon semileptonic
	decay data\cite{Dqextract}{}.
	It has been recently recognized\cite{Kamionkowski} that the new
	preliminary
	SMC measurements\cite{SMC2} of the spin structure function of
	the proton at 	$Q^2 =10.3$ GeV$^2$
	may have dramatic implications for calculations of the spin-dependent
	neutralino-nucleus scattering cross section.
	The values of $\Delta q$ extracted from these new data
	in comparison with previous EMC\cite{EMC} data are much closer
        to $SU(3)$ na\"{\i}ve quark model (NQM)
        predictions\cite{Witt,NQM}{}.
        This  gives rise to small enhancement of the spin-dependent
	cross section for
	nuclei with an unpaired proton and a strong (by a factor of about 30)
	suppression for nuclei with an unpaired neutron.
        In view of this
        we use in the analysis $\Delta q$ values
	extracted both from the EMC\cite{EMC} and from SMC\cite{SMC2} data.

        The other nuclear structure parameters $f_s$ and $\hat f$
        in formula  \rf{rel1} are defined as follows:
\ba{Scal} 
   <p(n)|(m_{u} + m_{d})(\bar{u}u + \bar{d}d)|p(n)> &=& 2\hat f M_{p(n)}
           \bar \Psi \Psi, \\
\nn
   <p(n)|m_{s}\bar{s}s|p(n)> &=& f_{s} M_{p(n)}\bar \Psi \Psi.
\ea
        The values extracted from the data under certain theoretical
        assumptions are\cite{ChengGasser}{}:
        \ba{f}
	\hat{f} = 0.05\ \ \ \ \ \ \mbox{and}\ \ \ \ \ \ f_{s} = 0.14.
        \ea
 	The strange quark contribution $f_{s}$ is known to be uncertain
        to about a factor of 2. Therefore we take its values in the analysis
        within the interval $ 0.07 < f_{s} < 0.3$\cite{Hatsuda,ChengGasser}{}.

        The nuclear structure comes into play via the form factors
        ${\cal F}_{ij}(q^2), {\cal F}_{S}(q^2)$ in Eq. \rf{cs}.
        The spin-independent
        form factor ${\cal F}_{S}(q^2)$ can be represented as the normalized
        Fourier transform of a spherical nuclear ground state density
	distribution $\rho({\bf r})$
\ba{fourier}
{\cal F}_{S}(q^2) = \int d^3{\bf r} \rho({\bf r}) e^{i {\bf r q}}.
\ea
	In the analysis we use the standard Woods-Saxon inspired
	distribution\cite{E}{}. It leads to the forma factor
\be{formf}
{\cal F}_{S}(q^2) = 3\frac{j_1(q R_0)}{q R_0} e^{-\frac{1}{2} (qs)^2},
\ee
	where $R_0 = (R^2 - 5 s^2)^{1/2}$ and $s \approx 1$ fm are
	the radius and the thickness of a spherical nuclear surface
	respectively, $j_1$ is the spherical Bessel function of index 1.

	Spin-dependent form factors ${\cal F}_{ij}(q^2)$ are much more
	nuclear model dependent quantities.
        The last few years have seen a noticeable progress
	in detailed nuclear model calculations of these form factors.
	For many nuclei of interest in DM search  they have been calculated
	within the conventional shell model\cite{Ressell} and
	within an approach based on the theory of finite Fermi
	systems\cite{Nikolaev}{}.
	We use the simple minded parametrization for $q^2$ dependence of
	${\cal F}_{ij}(q^2)$ in the form of a Gaussian with the rms spin
	radius of the nucleus calculated in the harmonic well
	potential\cite{EF}{}.
	For our purposes this semi-empirical scheme is sufficient.

        An experimentally observable quantity is the differential event rate
        per unit mass of the target material. It reads
\be{drate1}
\frac{dR}{dE_r} = \preRate
    \int^{v_{max}}_{v_{min}} dv f(v) v \dsdq2 (v, E_r)
\ee
	Here $f(v)$ is the velocity distribution of neutralinos in the
	earth frame which is usually assumed to be the Maxwellian
	distribution in the galactic frame.
	$v_{max} = v_{esc} \approx 600 km/s$ is the escape velocity at
	the sun position;
	$v_{min} = \left(M_A E_r/2 M_{red}^2\right)^{1/2}$ with
	$M_A$ and $M_{red}$ being the mass of nucleus $A$ and the reduced
	mass of the neutralino-nucleus system respectively. Note that
	$ q^2 = 2 M_A E_r$.

	The differential event rate is the most appropriate for comparing with
	the observable recoil spectrum and allows one to take properly into
	account spectral characteristics of a specific detector and to
	separate the background.
	However, for a more general theoretical discussion the event rate
	integrated over some domain of recoil energy is more useful and
	commonly employed for estimating the prospects for dark matter
	detection, ignoring experimental complications which may occur
	on the  way.
	Moreover, the integrated event rate is basically less sensitive
	to details of nuclear structure then the differential one \rf{drate1}.
	The $q^2$ shape of the form factors ${\cal F}_{ij}(q^2),
	{\cal F}_{S}(q^2)$ in Eq. \rf{cs} may essentially change from
	one nuclear model to another.
	Integration over some $q^2$ region reduces these variations
	drastically.

	Define the integral event rate as:
\be{rate1}
R(E_1, E_2) =
    \int^{E_{2}}_{E_{1}} \frac{dR}{dE_r}
\theta(E_{max} - E_r) dE_r
\ee
	Here $E_{max} = 2 M_{red}^2 v_{esc}^2/M_A$
	It is a common practice in theoretical papers to analyze the total
	event rate $R = R(0,\infty)$.
	For the realistic DM detector one should take into account a
	non-zero threshold energy $E_1 \geq E_{thr}$. In what follows we use
	the total event rate $R$ in a general discussion.
	The effect of a DM detector	threshold $E_{thr}$
	is analyzed later on in Sec. 2.

	Now let us address the question of possible constraints on SUSY models
	reachable in the direct DM search experiments.
	As seen from Eq. \rf{cs} there are just three parameters
	$a_{0,1}, c_0$ accumulating all SUSY model dependence via
	parameters $A_q, C_q$.
	Therefore, in the experiments discussed here the only three
	constraints on the combined SUSY model parameters
	$a_{0,1}, c_0$ are accessible.

	For getting a more transparent physical meaning of
	these constrains it is useful to adopt the approximation of
	the odd group model (OGM)\cite{EV}{}.
	It assumes for odd-even nuclei that a dominant contribution
	to the nuclear spin comes from the odd nucleon group.
	For the most of nuclei of interest for DM searches this approximation
	is fairly good.
	If so, one can write the following formula for the total
	event rate\cite{BKK}
\be{Rate1} 
   R_{p(n)} = \Bigl[\phi_{p(n)}\cdot a_{p(n)}^2
          + \phi_{0}\cdot c_0^2 \Bigr]
         \frac{\mbox{events}}{\mbox{kg}\cdot\mbox{day}}.
\ee
        for the case of proton(neutron) odd group nucleus.
        Here, $a_{p(n)} = a_0 \pm a_1$.
        The parameters $\phi_{i}$ depend on properties of the target nucleus
        as well as on the mass density and the average velocity of
        DM particles in the solar vicinity.

        The quantities $a_{p,n}$ and $c_0$ contain all
        dependence on the parameters of the effective Lagrangian \rf{Lagr}
        and do not depend on nuclear properties.
        This factorization leads us to the following conclusions.

        From Eq.\rf{Rate1} we see again that measuring the event rate $R$
        we can study just three special combinations of fundamental
        parameters $a_{p,n}$ and $c_0$.
        This is the only information about  fundamental parameters accessible
        in DM search experiments.
        R is a linear combination of the quantities $a_{p,n}$
        and $c_0$.
        To extract experimental limitations for each of them one should
        search for DM with different target nuclei.
        We can distinguish three categories of DM detectors
        with respect to their sensitivity to $a_{p,n}$ and $c_0$.
        These are detectors built of spin-non-zero target nuclei with
        an odd proton(neutron) group probing
        a linear combination  $a_{p(n)}$ and $c_{0}$ and spin-zero
        target nuclei sensitive only to the scalar part $c_0$ of
        the neutralino-nucleus interaction.

        We would like to stress again the following.
        To extract all possible information about SUSY-model parameters
        from direct DM search one should have three above mentioned types
        of DM detectors.
        No other information can be obtained from the direct
        DM search experiments.
        Different detectors can only improve the data on the three
        above-defined groups of SUSY-model parameters.

\begin{figure}[t]
\begin{picture}(6,3.7)
\put(0.4,1.2){\includegraphics{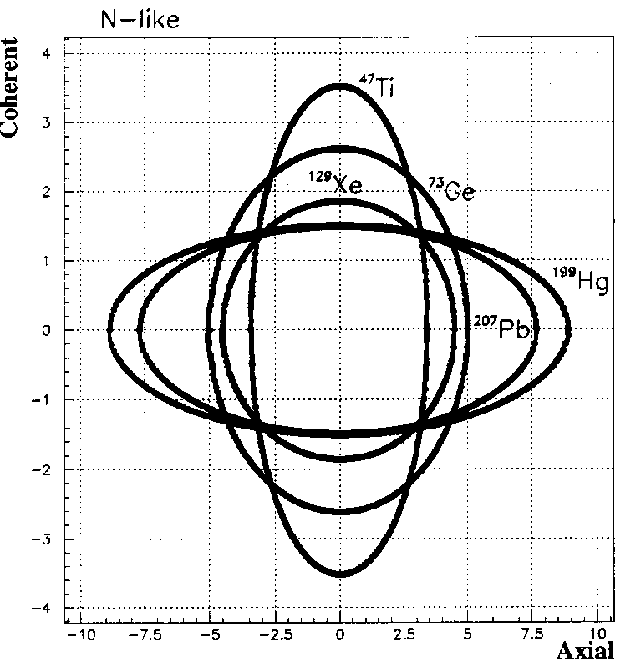}}
\put(3.2,1.1){\includegraphics{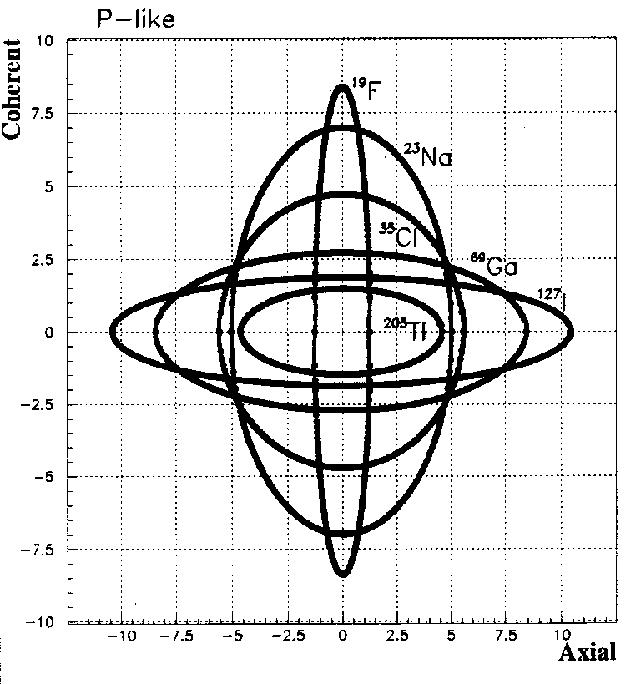}}   
\put(0.5,0.5)
               { \begin{minipage}[c]{5.in}
                FIG. 1. \
                Exclusion plots for the effective SUSY parameters
                $a_{p(n)}$ (Axial), $c_0$ (Coherent)
                for nuclei with odd  neutron (left panel) and
                proton (right panel) groups respectively.
	        \end{minipage}
}
\end{picture}
\end{figure}

        Accessible constraints can be represented as experimental
        constraints to the effective parameters $a_{p,n}$
        and $c_0$ accumulating all SUSY model dependence of the event rate.
        According to formula \rf{Rate1} the corresponding exclusion
        plots in the $a_{p}-c_0$ and $a_{n}-c_0$ planes
        have a form of ellipses as presented in
        Fig.1 for nuclei with neutron and proton odd groups respectively.
        The plots correspond to a DM detector sensitivity
        $R>1 \frac{events}{kg\cdot day}$. It is easy to see that
        the same picture holds for any sensitivity. The only effect is
        the rescaling of the $a_{p(n)}-c_0$  axes.
        These plots allow one to assess which isotope gives more
        restrictive constraints at the same detector sensitivity
        $R\geq R_{det}$. For instance, it is seen that among p-odd
        nuclei $^{205}$Tl has potentially better prospects
        as a target material for DM detectors than
        $^{23}$Na, $^{35}$Cl, $^{69}$Ga, $^{127}$I.
        Of course, one should remember that such a criterion is very
        superficial because it does not take into account
        limitations imposed by specific conditions under which an isotope
        can be used in a realistic DM detector.


\section{SUSY-model Predictions}

        In order to obtain quantitative predictions for the DM detection
        event rate one should calculate
        the parameters ${\cal A}_{q}$, and ${\cal C}_{q}$ of
        the effective Lagrangian \rf{Lagr} in the specific SUSY model.
        We  follow the MSSM with the GUT unification conditions for
        gauge coupling constants and for soft SUSY breaking parameters.
        This model is specified by the standard
	$SU(3)\times SU(2)\times U(1)$ gauge couplings as well as
        by the low-energy superpotential and soft SUSY breaking
        terms.

        The effective low-energy superpotential is:
\be{SUPERPOTENTIAL} 
         \tilde{W} = \sum_{generations}
          (h_L {\hat{H}}_1 {\hat{L}}{\hat{E}} +
	h_D {\hat{H}}_1 {\hat{Q}}{\hat{D}} -
	h_U {\hat{H}}_2 {\hat{Q}}{\hat{U}}) - \mu {\hat{H}}_1 {\hat{H}}_2.
\ee
	Here $\hat{L},\  \hat{E}$ are lepton doublets and singlets; $\hat{Q}$
	are quark doublets,
	$\hat{U},\  \hat{D}$ are {\em up} and {\em down} quark singlets;
        $\hat{H}_1$ and $\hat{H}_2$ are the Higgs doublets
	with a weak hypercharge $Y=-1,+1$, respectively.

        The effect of "soft" supersymmetry breaking can be parametrized
	at the Fermi scale as a part of the scalar potential:
\ba{V_soft} 
V_{soft}= \sum_{i=scalars}^{}  m_{i}^{2} |\phi_i|^2 +
h_L A_L H_1 \tilde L \tilde E + h_D A_D H_1 \tilde Q \tilde D
 - h_U A_U H_2 \tilde Q \tilde U  -&& \\
\nn
- (\mu B H_1 H _2 + \mbox{ h.c.})  &&
\ea
	and a "soft"  gaugino mass term
\be{M_soft} 
{\cal L}_{FM}\  = \ - \frac{1}{2}\left[M_{1}^{} \tilde B \tilde B +
 M_{2}^{} \tilde W^k \tilde W^k  + M_{3}^{} \tilde g^a \tilde g^a\right]
 -   \mbox{ h.c.}
\ee
	As usual, $M_{1,2,3}$ are the masses of the
	$SU(3)\times SU(2)\times U(1)$
        gauginos $\tilde g, \tilde W, \tilde B$.
        $m_i = \{ m_{L}, m_{E}, m_{Q}, m_{U}, m_{D}, m_{H_1}, m_{H_2}\}$
        are the mass parameters of scalar fields.

        To reduce the number of free parameters we use
	the following unification conditions at the GUT scale $M_X$:
\be{boundary1}  
 A_U = A_D = A_L = A_{0},
\ee
\be{boundary2}  
 m_{L}=m_{E}=m_{Q}=m_{U}=m_{D}=m_{0},
\ee
\be{boundary3}  
 M_{1}^{} = M_{2}^{} = M_{3}^{} = m_{1/2}^{},
\ee
\be{boundary4} 
g_{1}^{}(M_X) = g_{2}^{}(M_X) = g_{3}^{}(M_X) =  g_{GUT}^{},
\ee
        where $g_{3}^{}, g_{2}^{}, g_{1}^{}$ are the
	$SU(3)\times SU(2)\times U(1)$ gauge coupling constants
	equal to $g_{GUT}^{}$  at the unification scale $M_X$.

	At the Fermi scale $Q\sim M_W$ these parameters
        can be evaluated on the basis of the renormalization group equations
        (RGE)\cite{RGE}{}.
        Equation \rf{boundary3} implies at $Q\sim M_W$
\be{M1_M2}
M_1 = \frac{5}{3} \tan^2\theta_W \cdot M_2, \ \ \ M_2 = 0.3 m_{\tilde g}.
\ee
Here $m_{\tilde g} = M_3$ is the gluino mass.
	One can see from \rf{boundary1}-\rf{boundary4} that we do not exploit
	the complete set of GUT unification conditions for the soft
        supersymmetry breaking parameters, which leads to the supergravity
	scenario with radiative electroweak gauge symmetry breaking.
	Specifically, we do not  unify Higgs soft masses
        $m_{H_1}, m_{H_2}$ with the others in Eq. \rf{boundary2}.
	Otherwise strong correlations in the supersymmetric particle
	spectrum would emerge, essentially
        attaching the analysis to a particular supersymmetric scenario.

        We analyze the Higgs sector of the MSSM at the 1-loop
        level\cite{1loop} with taking into account
	$\tilde{t}_{L}-\tilde{t}_{R}$, $\tilde{b}_{L}-\tilde{b}_{R}$ mixing
	between the third-generation squarks.
        Diagonalization of the Higgs mass matrix leads to three neutral
        mass-eigenstates: two $CP$-even states, $h$, $H$, with
	the masses $m_h$, $m_H$ and the relevant mixing angle
	$\alpha_H$ as well as  one $CP$-odd state $A$ with the mass $m_A$.
        We take the mass $m_A$ as an independent free parameter.

        A complete list of the essential free parameters we use
	in the analysis is
\ba{param} 
\tan\beta, A_{0},\mu, M_2, m_A, m_0, m_t.
\ea
        The angle $\beta$ is defined by the vacuum expectation
        values of the neutral components of the Higgs fields:
        $\tan\beta = <H_{2}^{0}>/<H_{1}^{0}>$.

        We fix further for definiteness
        $m_t = 174$ GeV which corresponds to the CDF value of
        the top quark mass\cite{top}{}.

        There are four neutralino mass eigenstates $\chi_i$ in the MSSM
\be{admix}
\chi_i = {\cal N}_{i1} \tilde{W}^{3} +  {\cal N}_{i2}  \tilde{B} +
        {\cal N}_{i3} \tilde{H}_{2}^{0} + {\cal N}_{i4} \tilde{H}_{1}^{0}.
\ee
        which are linear combinations of zero charged gauginos
        $(\tilde{W}^{3}$, $\tilde{B}$ and Higgsinos
        $\tilde{H}_{2}^{0}$, $\tilde{H}_{1}^{0})$. The unitary matrix
        ${\cal N}$ rotates the neutralino $4\times 4$ mass matrix to
        the diagonal form. As usual, we denote the lightest neutralino
        $\chi_1$ as $\chi$.

        Having specified the model one can derive the effective
        Lagrangian $L_{eff}$ of low-energy neutralino-quark interactions.
        In the MSSM the first term of $L_{eff}$ in Eq. \rf{Lagr}
	is induced by the $Z$-boson
        and $\tilde{q}$ exchange whereas the second one is
        due to the Higgs boson and $\tilde{q}$ exchange.

        Our formulas for the coefficients ${\cal A}_q$ and ${\cal C}_q$
        of the effective Lagrangian take into account squark mixing
	$\tilde{q}_L-\tilde{q}_R$ and the contribution of
	both CP-even Higgs bosons $h, H$.
        As pointed out in Ref.\cite{Kam}{},
	the contribution of the heavier Higgs
        boson $H$  can be important in certain cases.
        At some values of the angles
	$\alpha_H, \beta$  and the neutralino composition
	coefficients ${\cal{N}}_{13},{\cal{N}}_{14}$ the contribution
	of the heavier Higgs boson $H$ to the coefficients ${\cal{C}}_q$
	can be larger than the contribution of the lightest Higgs boson $h$.
        The above formulas coincide with the relevant formulas
        in Ref.\cite{Drees} neglecting the terms $\sim 1/m_{\tilde q}^4$
	and higher.
        As stated in Sec. 2, we adopt the approximate treatment proposed
        in Ref.\cite{Drees}{}. It allows one to take into account these terms
	"effectively" by introducing an "effective"  stop quark $\tilde t$
        propagator.
        In the limit $\theta_q \rightarrow 0$,
        where $\theta_q$  is the $\tilde{q}_L - \tilde{q}_R$ mixing angle,
        our formulas agree with Ref.\cite{EF} except for the relative sign
        between the $Z$ and $\tilde{q}$ exchange terms in
	the coefficients ${\cal{A}}_q$  and up to the overall  sign in
        the coefficients ${\cal{C}}_q$. These errors in Ref.\cite{EF} were
        also observed in Ref.\cite{Drees}{}.
        Now we are ready to calculate the event rate of neutralino nucleus
        scattering.

	In our numerical analysis we scan the MSSM parameter
	space within a broad domain
\ba{domain}
 20\ GeV < M_2 &<& 1\ TeV,\ \ \ \ \  \ \ \ \ \ \ \ \ \ \ \ \ \ \
		                      |\mu| < 1\ TeV, \\ \nn
 1 < \tan\!\beta &<& 50, \ \ \ \ \ \ \ \ \ \ \ \ \ \ \ \
                          \ \ \ \ \ \ |A_0| < 1\ TeV,\\ \nn
 0 < m_0 &<& 1\ TeV, \ \ \ \ 50\ GeV < m_A < 1\ TeV.
\ea

	In the region where $\tan\!\beta \gsim 35$ the top Yukawa dominance
	approximation is not applicable in the RGE.
	Therefore, we use procedure developed in Ref.\cite{WBK}
	which takes into account the bottom and tau Yukawa couplings as well.

	Further limitations on  the parameters space are imposed
        by the known experimental lower bounds on supersymmetric
        particle and Higgs boson masses  from  LEP and Tevatron
        measurements.

	The neutralino relic density $\Omega_{\chi}$ is also under
	control in our analysis. We calculate it following the standard
        procedure on the basis of the approximate
	formula\cite{EHNOS,Relic,GKT}{}:
\ba{Omega}
\Omega_{\chi}h_0^2 =  2.13\times 10^{-11} \left(\frac{T_{\chi}}
	{T_{\gamma}}\right)^3 \left(\frac{T_{\gamma}}{2.7 K}\right)^3
	\cdot N_{F}^{1/2} \left(\frac{GeV^{-2}}{a x_F + b x_F^2/2}\right).
\ea
	Here $T_{\gamma}$ is the present day photon temperature,
	$T_{\chi}/T_{\gamma}$ is the reheating factor,
	$x_F = T_F/M_{\chi} \approx 1/20$, $T_F$ is
	the neutralino freeze-out temperature and $N_F$ is the total
	number of relativistic degrees of freedom at $T_F$.
	The coefficients $a, b$ are determined from the expansion
       \be{ts}
        <\sigma_{ann}v > \approx a  + b x
	\ee
	of the thermally-averaged cross section
	$<\sigma_{ann}v>$ of neutralino annihilation. We use an
        approximate treatment ignoring complications which occur
        when expansion \rf{ts} fails\cite{Seckel}{}.
	We take into account all possible channels of the $\chi-\chi$
        annihilation.

	It is well known that cosmologically acceptable neutralinos
	should produce a relic density  in the interval
\be{OmegaBound}
	0.025 < \Omega_{\chi}h_0^2 < 1.
\ee
	In this case neutralinos do not overclose the universe
	and account for a significant fraction 	of the halo DM.

\begin{figure}[t]
\begin{picture}(6,3.5)
\put(-0.0,1.0){\includegraphics{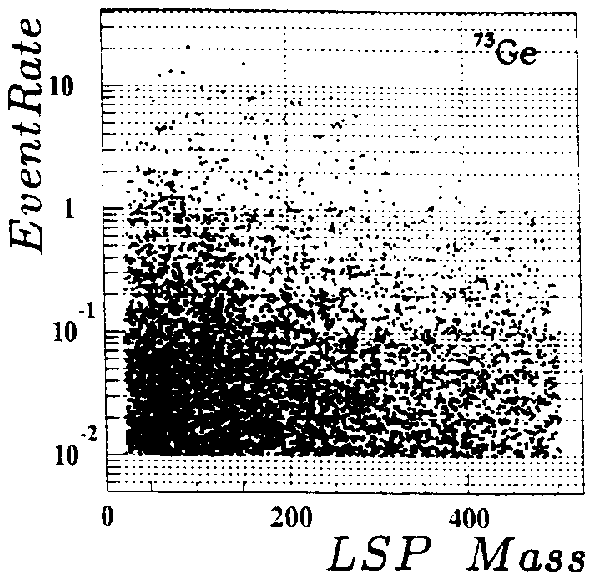}}
\put( 2.6,1.2){\includegraphics{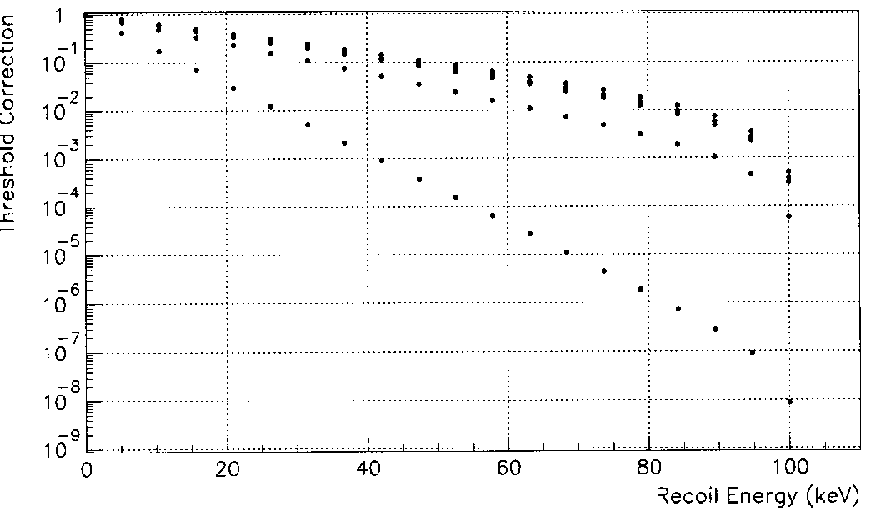}}
\put(0.5,0.5)
{ \begin{minipage}[c]{5.in}
                FIG. 2. \
                Scatter plots of the total event rate $R$
                vs the neutralino (LSP) mass $M_{\chi}$
		(left panel)
                and
                the threshold correction factor $\nu$ vs the
                threshold recoil energy
                $E_{thr}$
                (right panel).
                Three curves correspond to
                $M_{\chi}$ = 30, 100, 200 GeV
                from the bottom to the top. All plots for $^{73}$Ge.
\end{minipage} }
\end{picture}
\end{figure}

        As an example of our event rate calculations\cite{BKK1} we show
        in Fig.2(left)
        the total event rate $R$ for $^{73}$Ge.
        The scatter plot was obtained by random point generation in the MSSM
        parameter space with the constraints discussed above.
        More detailed presentation of these results including other isotopes
        is given in Ref.\cite{BKK1}{}.

        A non-zero threshold energy $E_r$ for the case of a realistic DM
        detector may essentially modify pictures displayed in
	Fig.2(left).
        To quantify the effect of a non-zero detector threshold we
        introduce the ratio
\be{thresh}
\nu(E_{thr}) = \frac{R(E_{thr}, \infty)}{R(0, \infty)}
\ee
	In Fig.2(right)
        this threshold factor is plotted for the isotope $^{73}$Ge.
	It is seen that for the lighter neutralino
	$\nu(E_{thr})$ falls faster than for the heavier one.
        One can easily understand this dependence noticing that
        a sizable $R$ can be obtained for $E_{thr} < 10^{-6} M_{\chi}$.
	The latter corresponds to the mean kinetic energy of the DM
	neutralino.

        Combining plots Fig.2(left) and Fig.2(right)
	we can estimate the maximal value of the event rate $R$(H-M)
	integrated from $E_{thr} = 48$ KeV which corresponds to the
	threshold recoil energy of the Heidelberg-Moscow germanium
	detector\cite{Klapdor}{}.
        As seen for $M_{\chi} \approx 200$ GeV, it approaches
	$R(H-M) = 0.2\frac{events}{kg\cdot day}$ comparable with
        the sensitivity quoted by this collaboration\cite{HM}{}.
	Therefore, one may expect this experiment to provide in the near
	future the DM constraints on the MSSM parameter space.
        Detailed analysis\cite{BKK1} leads us to the conclusion
        that in the similar
	position is another germanium experiment by the Caltech-PSI-Neuchatel
	collaboration\cite{CPN}{}.
        Other DM experiments\cite{TAUP} are
	more distant from the reach of the allowed part of
        the MSSM parameter space as yet.

        Note that in the analysis we ignore possible rescaling of the local
        neutralino density $\rho$ which may occur in the region of the
        MSSM parameter space where $\Omega h^2< 0.05$\cite{Gelm}{}.
        This effect, if it took place, could modify the predictions
        for the event rate $R$\cite{Bot}{}.
        We assume that neutralinos constitute a dominant component of
        the DM halo of our galaxy with a density
        $\rho$ = 0.3 GeV$\cdot$cm$^{-3}$ in the solar vicinity.

\section{The role of target nucleus spin}

       To study the role of nuclear spin in elastic $\chi$-nucleus
        scattering we consider the ratio
\be{eta} 
        \eta = R_{sd}/R_{si}
\ee
        characterizing the relative contribution of spin-dependent and
        spin-independent interactions. Here $R_{sd}$ and $R_{si}$
        are the spin-dependent and spin-independent parts of
        the total event rate $R$ respectively.
        The quantity $\eta + 1$ determines the expected
        relative sensitivity of DM detectors with spin-non-zero
        (J $\neq$ 0) to those with spin-zero \mbox{(J = 0)} nuclei
        as target material, if their atomic masses are close in value.
        If $\eta < 1$, then detectors with spin-non-zero and
        spin-zero target materials have approximately equal sensitivities
        to the DM signal, otherwise if $\eta > 1$, the spin-non-zero
        detectors are more sensitive than the spin-zero ones.

        Let us consider separately the dependence of $\eta$ on the nuclear
        structure parameters and on the parameters of neutralino-quark
	interactions determined in a specific SUSY model.
        Within our approximations one can write\cite{BKK1}{}:
\be{separat} 
   \eta = \eta_{A}^{} \eta_{susy}^{p(n)},
\ee
	The factorization \rf{separat} of the nuclear structure
	$\eta_A$ from
	the supersymmetric part of the neutralino-nucleus interaction
	$\eta_{susy}$ is essentially based on the assumption of the odd
	group model\cite{EV} about a negligible contribution of the even
	nucleon group to the total nuclear spin.
        $\eta_{A}$ is a factor depending on the properties of the
        nucleus $A$, while $\eta_{susy}^{p(n)}$ is defined by
	the SUSY-model which
        specifies the neutralino composition and the interactions with matter.
        The SUSY-factor also depends on the nucleon matrix element parameters
	\rf{Spin}, \rf{Scal} and on the shell-model class
        to which nucleus $A$ belongs,
        being $\eta_{SUSY}^{n}$  for the shell-model "neutron"
        (${}^{3}$He, ${}^{29}$Si, ${}^{73}$Ge, ${}^{129,131}$Xe,...)
	and $\eta_{SUSY}^{p}$ for
        the shell-model "proton" (${}^{19}$F, ${}^{23}$Na, ${}^{35}$Cl,
	${}^{127}$I, ${}^{205}$Tl,...).

\begin{figure}[t]
\begin{picture}(6,4)
\put(0.0,0.0){\includegraphics{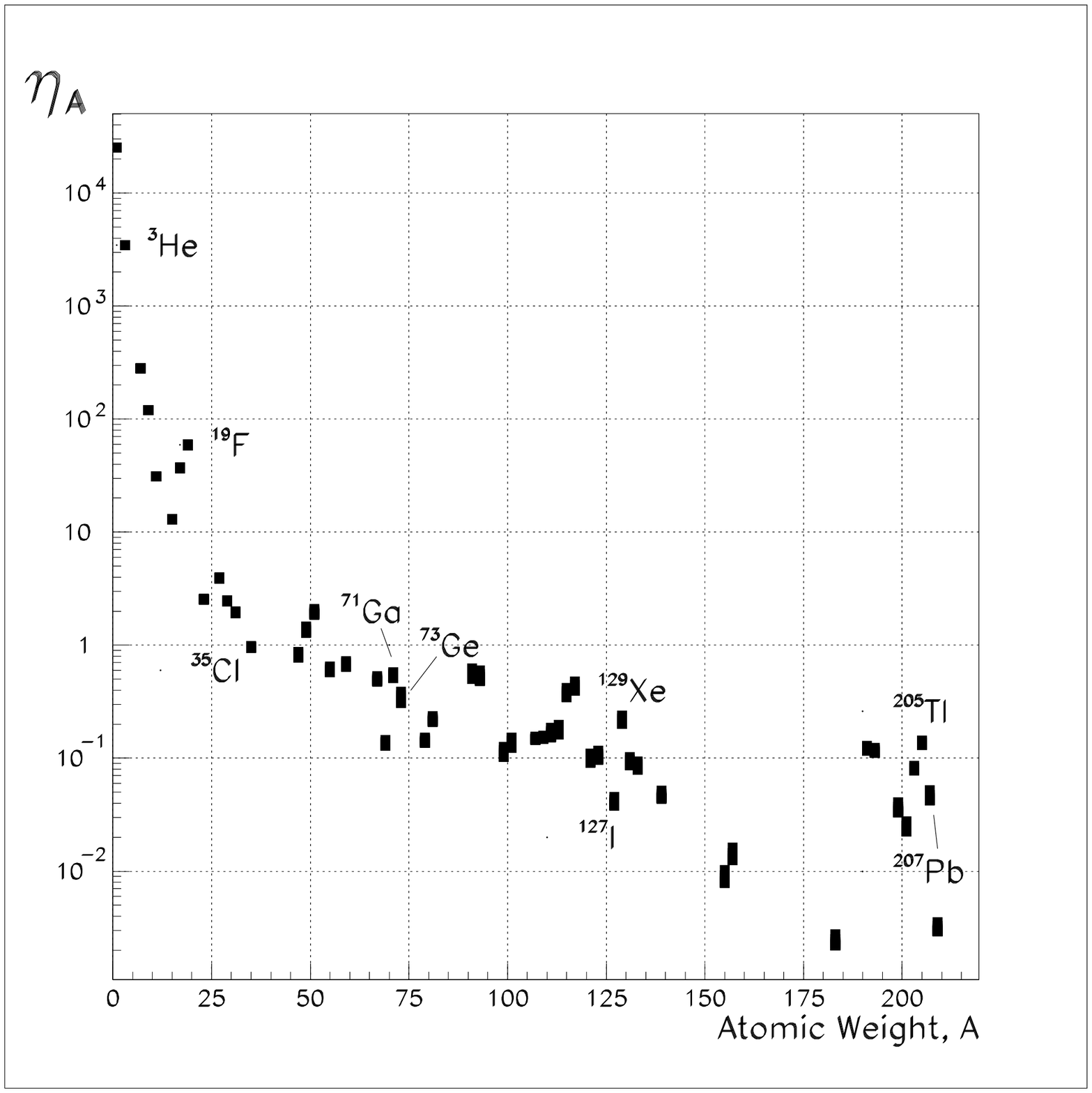} }
\put(0.9,0.2)
{ \begin{minipage}[c]{4.0in}
                FIG. 3. \
                The nuclear factor $\eta_{A}$ vs
        	the atomic weight $A$ for nuclei with non-zero spin.
\end{minipage} }
\end{picture}
\end{figure}

        Fig.3 shows the nuclear factor $\eta_{A}$ versus
        the atomic weight $A$\cite{BKK1}{}.
        The height of the symbols in the picture represents the variation
	of the ratio $\eta_{A}$
	within the explored interval of the neutralino mass of
	$20$ GeV $\leq M_{\chi} \leq  500$ GeV.

        It follows from Fig.3 that $\eta_{A} < 1$ for $A > 50$.
        Thus at $A > 50$ there is {\em no nuclear structure enhancement} of
        the spin-dependent event rate as compared to the spin-independent one.

        The next step is calculation of the SUSY-factor $\eta_{susy}^{p(n)}$
        within the MSSM. We have performed  numerical analysis of
        the MSSM parameter space as described in the previous section.
        The following absolute upper bound for the SUSY-factor in
        Eq. \rf{separat} was found:
\be{main}  
\eta_{susy}\  \ \lsim \ \  2,
\ee
        in the subdomain of the parameter space where the total event rate is
        $R\ \gsim\ 0.01 \frac{event}{kg\cdot day}$.

\begin{figure}[t]
\begin{picture}(6,3.5)
\put(0.5,1.0){\includegraphics{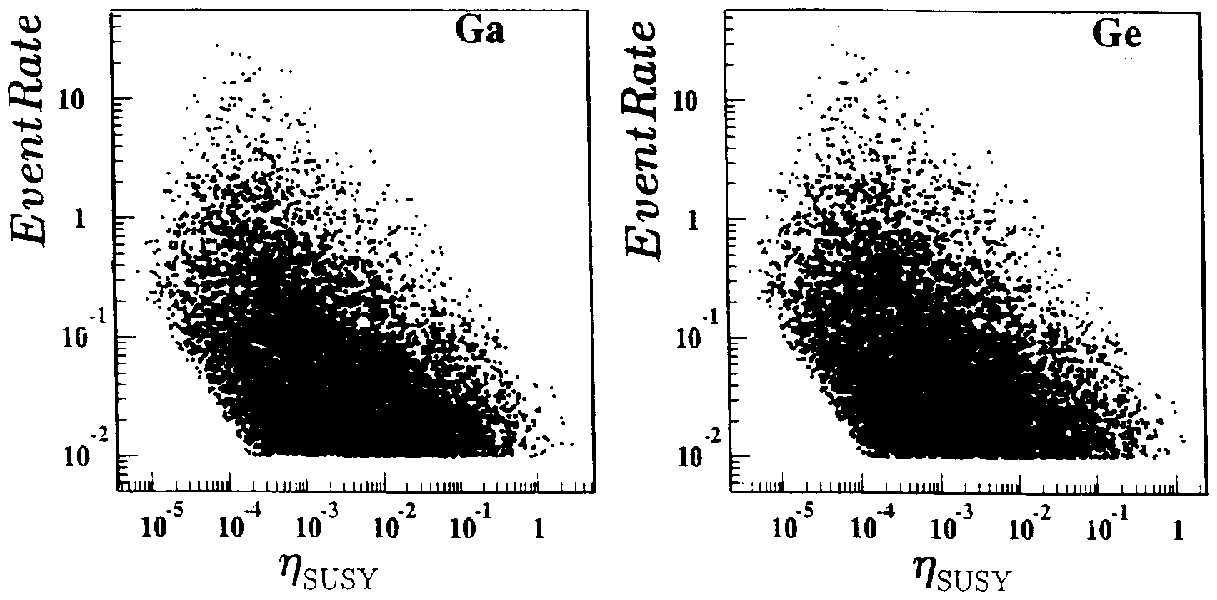}}
\put(0.5,0.5)
               { \begin{minipage}[c]{5.in}
                FIG. 4. \
                Scatter plots of the total event rate $R$
                vs  the ratio $\eta_{susy}$.
                Two representative nuclei with an unpaired proton
                ($^{71}$Ga) and an unpaired neutron
                ($^{73}$Ge) are presented.
	        \end{minipage}
}
\end{picture}
\end{figure}

	Fig.4 shows the distribution of the points in the $R-\eta_{susy}$
        plane. Plots are given for two representative nuclei with an unpaired
        proton (p-like), ${}^{71}$Ga, and with an unpaired neutron (n-like),
        ${}^{73}$Ge. The nuclei are taken for convenience near the point
        $A = 50$ (see Fig.3).
        For heavier nuclei we have obtained
	basically the same picture and our further conclusions correspond
        to all nuclei with $A > 50$. One can see the above quoted \rf{main}
        upper bound $\eta_{susy}\ \lsim\  2$ for both cases
        presented in Fig.4.

        Now we may combine the bound \rf{main} with the values of
        the nuclear factor  $\eta_A$  represented in Fig.3.
        Then we obtain the conservative estimate:
\be{main1} 
  \eta = R_{sd}/R_{si} =
   \eta_A^{} \eta_{susy}^{p(n)}{} \ \lsim 1.6 \
        \mbox{ ~for nuclei with~} A > 50
\ee
        at a detector sensitivity up to
	$R > 0.01 \frac{ events }{  kg \cdot day }$.
        However, as is seen in Fig.4, the majority of points
        generated in the domain \rf{domain} of the MSSM parameter space are
        concentrated at $\eta \ \leq \ 1$.
        The tendency is that at higher sensitivities (lower $R$ accessible)
        we get $\eta \leq 1$ for heavier nuclei and vice versa.

\begin{figure}[t]
\begin{picture}(6,2.8)
\put( 0.4,1.0){\includegraphics{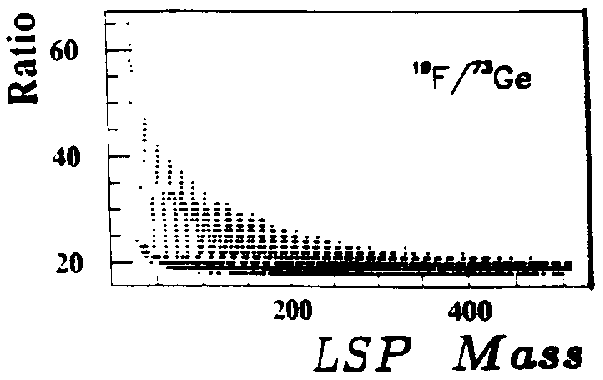}}
\put( 3.1,1.0){\includegraphics{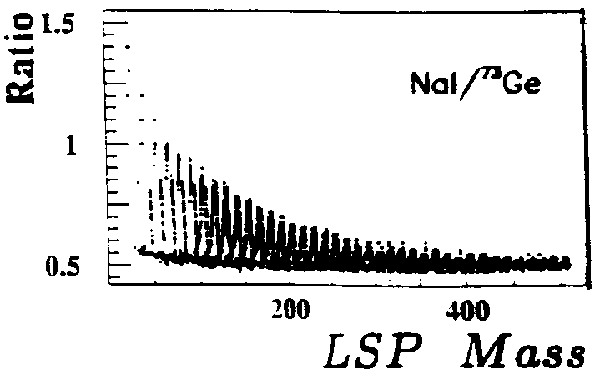}}
\put(0.5,0.5)
               { \begin{minipage}[c]{5.in}
                FIG. 5. \
 	        The ratio $r(A) = R_{sd}(A)/R_{sd}({}^{73}\mbox{Ge})$
		of the spin-dependent event rate $R_{sd}$
                for nuclei ${}^{19}$F and NaI to
		the spin-dependent event rate $R_{sd}({}^{73}\mbox{Ge})$
                for ${}^{73}$Ge.
\end{minipage}}
\end{picture}
\end{figure}

        In Fig.5 we also present
	plots of the ratio $r(A) = R_{sd}(A)/R_{sd}({}^{73}\mbox{Ge})$
	of the spin-dependent part $R_{sd}$ of the event rate
        for some target materials $(A)$ to that for ${}^{73}$Ge.
        This ratio for ${}^{129}$Xe is
        $r({}^{129}\mbox{Xe}) \approx 1.2$ being
        fairly independent of the neutralino mass $M_{\chi}$. As explained
	in Ref.\cite{BKK1} this is the case because
	both ${}^{73}$Ge and  ${}^{129}$Xe
        are nuclei with an unpaired neutron.

        It follows from our detailed analysis\cite{BKK1} and is illustrated
        by Fig.5 that the maximal values of
	the total event rate for NaI, ${}^{73}$Ge and ${}^{129}$Xe
	are typically the same while for CaF$_2$ they are
	lower by about a factor of 5.
        On the other hand, the sensitivity of CaF$_2$
	to the spin-dependent part of the neutralino-nucleus interaction
	is by about a factor of 10 larger than that of NaI, ${}^{73}$Ge and
	${}^{129}$Xe.  The last three materials have approximately
	an equal spin sensitivity.

\bigskip

\section{Conclusion}

        The central point of this report is that the operating Dark Matter
	detectors are in the position to probe the MSSM parameter space in
	the near future.
	We argued that there are the only three sorts of constraints
	attainable in direct DM searches.
	In this respect DM detectors fall into three different
	categories probing different combinations of SUSY model parameters.
	They are detectors built of target nuclei with zero spin and with
	non-zero spin having odd proton, odd neutron groups.

        It was pointed out that for sufficiently
	heavy nuclei with atomic weights $A > 50$
        the spin-independent event rate $R_{si}$ is typically
	larger than
        the spin-dependent one $R_{sd}$ if low rate DM signals
        with total event rates
	$R =R_{sd} + R_{si} < 0.01 \frac{ events }{  kg \cdot day }$
	are ignored.
        This cut-off condition reflects the realistic sensitivities of
        the present and the near-future DM detectors.

        The main practical issue is that two different DM detectors with
        (J~=~0, A$_1$) and with (J~$\neq$~0, A$_2$) nuclei as a target
        material have equal chances  to discover DM events
        if  A$_1\sim$ A$_2 >$ 50.

        Another aspect of  the DM search is the investigation of the
        SUSY-model parameter space  from nonobservation of DM events.
        For this purpose experiments both with J~$=$~0 and  J~$\neq$~0
        nuclei are important.

	We have compared several examples of popular
	(see for instance \cite{TAUP} and references therein)
	materials with non-zero spin nuclei
	as a target in a  DM detector.
	We have not found an essential difference between NaI, ${}^{73}$Ge
	and ${}^{129}$Xe as a target material for DM detectors from
	the point of view of their total and spin sensitivity.
        We expect these materials to have better prospects as compared
        with  CaF$_2$  for discovering DM events.
	The former materials have a total
	event rate by about a factor of 5 larger than the latter one.
        On the other hand, CaF$_2$ can give a more stringent
	constraint on the spin-dependent part of the event rate, having
	a spin sensitivity by about a factor of 10 larger than
	NaI, ${}^{73}$Ge and ${}^{129}$Xe.

	Estimating prospects for various isotopes as target materials of
	DM detectors one may use the criterion given in Sec. 1 in the
	form of exclusion plots (Fig.1) for the effective SUSY model
	parameters. These plots characterize sensitivity of an isotope
	to the DM event ignoring possible complications
	of their employment in a detector.

	In conclusion we would like to stress that the efforts to improve
	existing and to construct new DM detectors are
	justified by the expectation that these detectors will be able
	to tackle supersymmetry from the side unreachable for
	the accelerator experiments. Therefore, these two types of
	experiments are complementary and should be considered as
	equally important in searching for supersymmetry.


\end{document}